\documentclass[10pt]{article}

\usepackage[english]{babel}

\usepackage[a4paper,top=2cm,bottom=2cm,left=3cm,right=3cm,marginparwidth=1.75cm]{geometry}

\usepackage{amsmath,mathrsfs,latexsym}
\usepackage{graphicx}
\usepackage{authblk}
\usepackage[colorlinks=true, allcolors=blue]{hyperref}

\title{\bf From Chern-Tenenblat to Jackiw-Teitelboim via sine-Gordon}
\author{Jeff Murugan}
\affil{\it The Laboratory for Quantum Gravity \& Strings,
Department of Mathematics \& Applied Mathematics,
 University of Cape Town, 
 Cape Town, South Africa}
\begin{document}
\maketitle

\begin{abstract}
\noindent
The study of 2-dimensional surfaces of constant curvature constitutes a beautiful branch of geometry with well-documented ties to the mathematical physics of integrable systems. A lesser known, but equally fascinating, fact is its connection to 2-dimensional gravity; specifically Jackiw-Teitelboim (JT) gravity, where the connection manifests through a coordinate choice that roughly speaking re-casts the gravitational field equations as the sine-Gordon equation. In this language many well-known results, such as the JT-gravity black hole and its properties, were understood in terms of sine-Gordon solitons and their properties. In this brief note, we revisit these ideas in the context of some of the recent exciting developments in JT-gravity and, more generally, low-dimensional quantum gravity and speculate on how some of these new ideas may be similarly understood. 
\end{abstract}

\section{Introduction}
The marriage between geometry and physics has  been a particularly fruitful one, and the name of S.S. Chern is instantly recognised as one of the principle architects at this interface. While his development of the theory of characteristic classes is familiar to most graduate students of theoretical physics by its application to topological quantum field theory, his contributions to nonlinear analysis are perhaps less so. This line of work is best exemplified in the seminal 1986 work of Chern and Tenenblat \cite{Chern86}, in which they introduced and studied in painstaking detail, a class of partial differential equations describing pseudospherical surfaces of the type pictured in Fig.1. To translate between these two descriptions, Chern and Tenenblat noted that the PDE
\begin{eqnarray}\label{pde}
 F\left(t,x,\phi,\frac{\partial\phi}{\partial t},\frac{\partial\phi}{\partial x},\ldots,\frac{\partial^{l}\partial^{k-l}\phi}{\partial t^{l}\partial x^{k-l}}\right) = 0\,,
\end{eqnarray}
describes a pseudospherical surface $\mathscr{M}$ if 1-forms $\omega_{i} =f_{i1}\,\mathrm{dx} + f_{i2}\,\mathrm{dt}$ can be found, with $i=1,2,3$ and smooth function $f_{ij}\left(t,x,\phi,\partial_{t}\phi,\partial_{x}\phi,\ldots\right)$, such that the associated {\it structure equations}
\begin{eqnarray}\label{structure}
   \mathrm{d}\omega_{i} = \frac{1}{2}\epsilon_{ijk}\,\omega_{j}\wedge \omega_{j}\,
\end{eqnarray}
hold if $\phi$ is a solution of \eqref{pde}. Conversely, given a solution of \eqref{pde} we can always construct a pseudosphere with metric $ds^{2} = \left(\omega_{1}\right)^{2} + \left(\omega_{2}\right)^{2}$, constant negative Gaussian curvature (which we can always normalize to $-1$), and for which we can identify $\omega_{3}$ as the Levi-Civita connection 1-form.\\

\noindent
Of course, not all differential equations are created equal. One of particular interest, both to Chern and Tenenblat in their original work, as well as to us in this article, is the {\it sine-Gordon} equation. To see how it arises in this construction, choose 1-forms
\begin{eqnarray}
   \omega_{1} = \cos\frac{\phi}{2}\left(\mathrm{dx} + \mathrm{dt}\right)\,;\quad
   \omega_{2} = 
   \sin\frac{\phi}{2}\left(\mathrm{dx} - \mathrm{dt}\right)\,;\quad
   \omega_{3} = 
   \frac{1}{2}\left(\partial_{x}\phi\,\mathrm{dx} - \partial_{t}\phi\,\mathrm{dt}\right)
\end{eqnarray}
so that \eqref{structure} holds iff the scalar satisfies 
\begin{eqnarray}\label{sG}
   \Box\phi(t,x) = \sin\phi(t,x)\,.
\end{eqnarray}
This is by no means a unique choice of 1-forms that produce the sine-Gordon equation. In fact, Chern and Tenenblat themselves use 
\begin{eqnarray}
   \omega_{1} = \frac{1}{\eta}\sin\phi\,\mathrm{dt}\,;\quad
   \omega_{2} = 
   \eta\,\mathrm{dx} + \frac{1}{\eta}\cos\phi\,\mathrm{dt}\,;\quad
   \omega_{3} = 
   \partial_{x}\phi\,\mathrm{dx}
\end{eqnarray}
as a one-parameter family of 1-forms that all yield the sine-Gordon equation for $\phi(t,x)$. Different choices of $\omega_{i}$ yield, alternatively, the KdV, modified-KdV, Burgers and sinh-Gordon equations, among others. In addition to being prototypical examples of {\it integrable} equations, each of these 2-dimensional classical field theories also possess a rich spectrum of solitonic solutions.\\

\noindent 
The connection to 2-dimensional gravity is equally fascinating. Certainly, as everyone knows, Einstein gravity in $(2+1)$-dimensions is (dynamically) trivial. However it is a (slightly) lesser-known fact that Einstein gravity in $(1+1)$-dimensional is {\it not even trivial}. This is because in two spacetime dimensions the Einstein tensor $G_{\mu\nu}\equiv R_{\mu\nu} - \frac{1}{2}g_{\mu\nu}R$ vanishes {\it identically}. To circumvent this rather boring state of affairs, Jackiw \cite{Jackiw:1984je} and, independently, Teitelboim \cite{Teitelboim:1983ux} made the suggestion to base the dynamical equations for {\it linear} gravity\footnote{Not to be confused with {\it linearized} gravity, which is something else entirely.} on the Reimann {\it scalar} instead, since in two dimensions, it is the simplest quantity that encodes all the local geometrical information about the spacetime.
The resulting dilaton gravity theory, known as Jackiw-Teitelboim (JT) gravity, not only possesses a great deal of symmetry ($SO(2,1)$ in $D=2$), cosmological \cite{Mann:1992yq} and black hole solutions \cite{Mann:1989gh} and even a gauge-theoretic formulation \cite{Isler:1989hq,Chamseddine:1989yz}, but - and this is the crux of this note - in fact {\it every} generic solution of pseudospherical equations of the form \eqref{pde} furnishes a classical solution of JT gravity \cite{Reyes:2006rf}. In this sense, it would seem that JT gravity is deeply connected to the theory of integrable systems.\\

\begin{figure}
\centering
\includegraphics[width=0.4\textwidth]{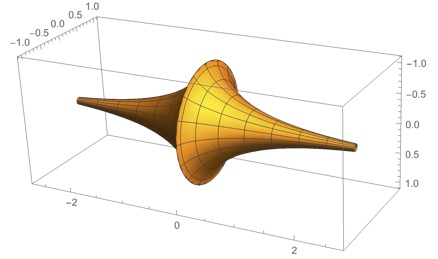}
\caption{\label{fig:Pseudo}A pseudospherical surface defined parametrically through $(x,y,z) = (\mathrm{sech}(u)\cos(v), \mathrm{sech}(u)\sin(v), u - \tanh(u))$.}
\end{figure}

\noindent
Spurred on by its relation to the Sachdev-Ye-Kitaev (SYK) model \cite{Maldacena:2016hyu,Kitaev:2017awl}, JT gravity is currently enjoying a dramatic renaissance; one that has produced a slew of results on, among others, gravity as an ensemble average \cite{Stanford:2019vob}, replica wormholes \cite{Penington:2019kki} and potential resolutions of the information loss paradox \cite{Hollowood:2020kvk}. A key feature of these new developments is the inclusion of the {\it boundary data} which encodes the Schwarzian dynamics \cite{Mertens:2018fds} and associated chaoticity properties of the theory. This is to be contrasted with the work carried out in the 1980's and 1990's which focused on the bulk dynamics. In this light then, it seems timeous to revisit the connection between JT gravity and integrable 2-dimensional models in the hope that there are  lessons to be learnt from the geometry of pseudo-spherical surfaces that may yet prove valuable for quantum gravity. Alas this article will not contain them. An exposition of a lecture presented at the 2021 Nankai Symposium,
it is instead two parts review (of the classical JT/sine-Gordon connection, following the treatment given in \cite{Gegenberg:1997ja,Gegenberg:1997ns,Gegenberg:1998rt}), one part wild speculation of how this connection may be used to understand some of the more contemporary questions asked of the 2-dimensional physics of JT gravity. The devil, as they say, is in the details, and these are left to forthcoming work \cite{Murugan:2022kki}.

\section{The sine-Gordon/JT connection}
The sine-Gordon equation \eqref{sG} is the paragon of a nonlinear PDE; integrable, possessing a rich spectrum of solitonic solutions and (it's quantum version is even) S-dual to the massive Thirring model via bosonization. Following the prescription above,
corresponding to the sine-Gordon equation \eqref{sG} is a pseudospherical Riemannian 2-manifold with metric
\begin{eqnarray}
   ds^{2} = \sin^{2}\frac{\phi}{2}\,dt^{2} + \cos^{2}\frac{\phi}{2}\,dx^{2}
\end{eqnarray}
Similarly, the (perhaps less familiar) {\it Euclidean} sine-Gordon equation maps to a {\it Lorentzian} pseudosphere according to
\begin{eqnarray}\label{Lor-pss}
   \left(\partial_{t}^{2}+\partial_{x}^{2}\right)\phi = \sin\phi\quad\longleftrightarrow\quad ds^{2} = -\sin^{2}\frac{\phi}{2}\,dt^{2} + \cos^{2}\frac{\phi}{2}\,dx^{2}\,.
\end{eqnarray}
Of course, which 2-manifold this is, depends on $\phi(t,x)$. In differential geometry, {\it B\"acklund transforms} map pseudospherical surfaces into each other via linear differential equations. Translated back into the language of PDEs, these act as solution-generating transforms that, starting with some (usually trivial) initial seed solution, generate arbitrarily many more. A familiar, albeit trivial, example of the B\"acklund transform is the first order {\it Cauchy-Riemann} equations,
\begin{eqnarray}
   \partial_{x}u = \partial_{y}v,\quad \partial_{y}u = -\partial_{x}v\,,
\end{eqnarray}
that relate the real and imaginary components of an analytic function $f(x,y) = u(x,y) + iv(x,y)$. Since each of these components themselves satisfy the second order Laplace equation $\Delta u(x,y) = \Delta v(x,y) = 0$, the Cauchy-Riemann equations act as an auto-B\"acklund transformation that maps the function $u(x,y)$ into its harmonic conjugate, $v(x,y)$. On the other hand, the B\"acklund transform
\begin{eqnarray}
   \partial_{x}v = \partial_{x}u + 2\kappa \exp\left(\frac{u+v}{2}\right),\quad \partial_{y}v = -\partial_{y}u - \frac{2}{\kappa} \exp\left(\frac{u-v}{2}\right)\,,
\end{eqnarray}
maps a harmonic function $v(x,y)$ into a solution of the (nonlinear) Liouville equation $\Delta u(x,y) = \exp(u(x,y))$ \cite{Rogers:1982qb}. Of course it is much simpler to solve the linear Laplace equation than the nonlinear Liouville equation. More to the point of this article, the linear system,
\begin{eqnarray}\label{sGBack}
   \partial_{X}v = \partial_{X}u + 2\kappa \sin\left(\frac{u+v}{2}\right),\quad \partial_{Y}v = -\partial_{Y}u + \frac{2}{\kappa} \sin\left(\frac{u-v}{2}\right)\,,
\end{eqnarray}
furnishes an auto-B\"acklund transformation, $\mathsf{B}_{\kappa}$ for the sine-Gordon equation\footnote{In the sense that both $u$ and $v$ each satisfy the sine-Gordon equation, with $X$ and $Y$ regarded as lightcone coordinates for the d'Alembertian operator.}, with parameter $\kappa$. To illustrate the power of \eqref{sGBack} as a solution-generating transform, let's start with the trivial solution of \eqref{sG}, $v = 0$, as a seed. Substituting into \eqref{sGBack}, integrating, and converting back to $(t,x)$ coordinates yields
\begin{eqnarray}\label{kink}
   u(t,x) = 4\tan^{-1}\left[\exp\left(\frac{x-x_{0}-v(t-t_{0})}{\sqrt{1-v^{2}}}\right)\right]\,.
\end{eqnarray}  
This is, of course just the sine-Gordon kink (or 1-soliton since it carries topological charge $Q=1$). More generally, if $\omega$ is any seed solution, and $\omega_{12\cdots N} \equiv \mathsf{B}_{\kappa_{N}}(\cdots \mathsf{B}_{\kappa_{2}}(\mathsf{B}_{\kappa_{1}}(\omega)))$, then the {\it Bianchi permutability theorem} \cite{Rogers:1982qb} gives an entirely algebraic construction of the $N$-soliton of the sine-Gordon equation, given the data $\{\omega,\omega_{1},\omega_{2},\ldots,\omega_{123\cdots N-1}\}$.\\

\noindent
To appreciate the connection to 2-dimensional gravity, consider the JT-gravity action functional,
\begin{eqnarray}
   I_{JT}[\tau,g_{\mu\nu}] = \frac{1}{16\pi G_{N}}\int_{\mathscr{M}}\!d^{2}x\,\sqrt{-g}\,\tau
   \left(R + \Lambda\right)\,, 
   \label{JT}
\end{eqnarray}
with (dimensionless) Newton's constant $G_{N}$, dilaton $\tau$, and cosmological constant $\Lambda$, which we will set to 2 in what follows. Solutions of the associated equations of motion, $R + 2 = 0$ and $(\nabla_{\mu}\nabla_{\nu} - g_{\mu\nu})\tau=0$, are manifestly constant curvature, and therefore maximally symmetric in the sense of admitting three independent Killing vectors $k^{\mu}_{(i)} = \epsilon^{\mu\nu}\partial_{\nu}\tau_{(i)} / \sqrt{-g}$, $i=1,2,3$. All such solutions must be locally diffeomorphic to $AdS_{2}$ however, because a solution is determined by both the metric $g_{\mu\nu}$ {\it and} the dilaton $\tau$, it may still be {\it globally} distinct from $AdS_{2}$. An example of this is the JT black hole,
\begin{eqnarray}\label{JTbh}
   ds^{2} = -\left(r^{2}-M\right)dt^{2} + \left(r^{2}-M\right)^{-1}dr^{2}\,,
\end{eqnarray}
where $M$ is constant and $r$ is linearly related to the dilaton. This solution, which can also be arrived at by dimensionally truncating a 3-dimensional BTZ black hole, clearly has a horizon at $r=\sqrt{M}$. While the BTZ black hole is singular at $r=0$, the same is not true of \eqref{JTbh}. However, there is a corresponding statement in JT gravity and that is that at $\tau=0$, $G_{N}$ blows up.\\

\noindent
To make the connection to pseudospherical surfaces, let's take as an ansatz for the metric the line element given in \eqref{Lor-pss}. Substituting into the JT action functional reduces it to the form,
\begin{eqnarray}
   I_{JT}[\tau,\phi] = \frac{1}{16\pi G_{N}}\int_{\mathscr{M}}\!d^{2}x\,\tau
   \left(\Delta\phi - \sin\phi\right)\,. 
\end{eqnarray}
Correspondingly, the equations of motion that arise from variation of $I_{JT}$ with respect to the dilaton and the embedding coordinate,$\phi$ are the Euclidean sine-Gordon equation, $\Delta\phi = \sin\phi$ and, respectively, the second order linear equation, $(\Delta - \cos\phi)\tau = 0$. An interesting observation to be made here is that, even though (one of) the equations of motion reduce to the sine-Gordon equation in this gauge, the JT action does not reduce to the sine-Gordon action. In a remarkable series of articles \cite{Gegenberg:1997ja,Gegenberg:1997ns,Gegenberg:1998rt}, Gegenberg and Kunstatter showed that the theory is nevertheless quite rich. Two points in particular should be singled out. The first is that taking $\phi$ to be the sine-Gordon kink \eqref{kink}, reduces the ansatz to
\begin{eqnarray}
   ds^{2} = -\mathrm{sech}^{2}\rho\,dt^{2} + \mathrm{tanh}^{2}\rho\,dx^{2}\,,
\end{eqnarray}
with $\rho\equiv (x-vt)/\sqrt{1-v^{2}}$ and $x_{0}=t_{0}=0$. Then, defining the new coordinates $(t',r')$ through
\begin{eqnarray}
   dt'^{2}\equiv dt^{2} - v\frac{\mathrm{\sqrt{1-v^{2}}\,sinh}^{2}\rho}{1 - v^{2}\mathrm{sinh}^{2}\rho}d\rho\,,\quad r'\equiv \sqrt{1-v^{2}}\,\mathrm{sech}\rho\,,
\end{eqnarray}
yields precisely the JT black hole \eqref{JTbh} if we identify the black hole mass $M$ with the (square of the) spectral parameter, $v^{2}$. The resulting metric has ADM energy $E = v^{2}/2G_{N}$, is singular at the location of the kink, $\rho=0$ and at $\rho=\infty$, and regular at the horizon $r' = v$. This construction is quite general; indeed the authors of \cite{Gegenberg:1998rt} go on to show that an $N$-soliton solution of the sine-Gordon equation (of which the kink is the 1-soliton) can be mapped to a black-hole metric of JT gravity with a unique mass.
\noindent
The second point, while more speculative is no less intriguing, and asks if the $N$-soliton construction may provide some insight into the nature of the Bekenstein-Hawking entropy of the JT black hole. Ignoring much of the exotica\footnote{These include, soliton-anti-soliton bound states, breathers etc.} of the sine-Gordon spectrum, the total rest energy of an $N$-soliton is given in terms of the sine-Gordon coupling $\beta$ by $E_{0} = N/\beta$. As argued earlier, the $N$-soliton state can be constructed from lower topological charge states. This is more than just a mathematical statement encoded in the Bianchi permutability theorem; it is a physical fact. The degeneracy of this state is then computed as the number of ways of writing $N$ as a sum of non-negative integers {\it i.e.} the {\it partition of $N$}, $p(N)$. While no exact closed form expression exists for $p(N)$, for large $N$ it can be approximated by the Hardy-Ramanujan formula, 
\begin{eqnarray}
   p(N) \sim \frac{1}{4N\sqrt{3}}\exp\left(\pi\sqrt{\frac{2N}{3}}\right)\,.
\end{eqnarray}
The corresponding entropy of the state $S\sim \log(p(N))\sim \pi\sqrt{N}$, matching the Bekenstein-Hawking entropy of the JT black hole \eqref{JTbh}, up to (admittedly important) factors of order one. As the authors themselves remark, this argument is rather coarse and in need of refinement. Still, it is hardly likely to be merely coincidental.

\section{Some speculative comments}
Having summarised, perhaps overly succinctly, what was known about the Chern-Tenenblat/sine-Gordon/Jackiw-Teitelboim connection around the turn of the last century, we now proceed to speculate as to how this may fit into the present day context in which, at least in the holographic quantum gravity community, JT gravity is enjoying somewhat of a deluge of attention. Much of this interest can be traced back to Kitaev's remarkable observation that the low-energy limit of JT gravity matches that of the SYK model of $N$ Majorana fermions interacting with quenched random $q$-body interactions \cite{Maldacena:2016hyu,Kitaev:2017awl}. Key to this observation is the Gibbons-Hawking boundary term
\begin{eqnarray}
   I_{bdy} = \frac{1}{8\pi G_{N}}\int_{\partial \mathscr{M}}\! \tau_{b}K\,,
\end{eqnarray}
that needs to be added to $I_{JT}$ in the holographic context, when the manifold $\mathscr{M}$ has a boundary $\mathscr{\partial M}$ in which the dual field theory resides. Here $K$ is the trace of the extrinsic curvature tensor and $\tau_{b} = \tau|_{\partial\mathscr{M}}$, is the value of the dilaton on the boundary. If, as in this case, $\partial\mathscr{M}$ is a co-dimension-1 curve parameterized by some `time' variable $u$ then $\mathscr{M}$ is cut out along the trajectory $(t(u),z(u))$. If, in addition, $T$ and $n$ denote the tangent vector field and unit normal to the boundary curve respectively, then 
\begin{eqnarray}
   K = -\frac{g(T,\nabla_{T}n)}{g(T,T)}= \frac{t'(t'^{2}+z'^{2} + zz'')-zz't''}{\left(t'^{2}+z'^{2}\right)^{3/2}}\,.
\end{eqnarray}
Implementing boundary conditions on the induced metric and dilaton of $ds|_{\partial\mathscr{M}} = du/\epsilon$, $\tau_{b} = \tau_{r}/\epsilon$ and sending $\epsilon\to 0$ allows the extrinsic curvature to be written in terms of the Schwarzian derivative as $K = 1 + \epsilon^{2}\mathsf{Sch}(t,u)$ to leading order in $\epsilon$. This defines the so-called {\it Schwarzian action} that encodes the leading order corrections to the conformal limit and, important for the identification with the dual quantum mechanics on the boundary, controls the chaos properties of the theory as diagnosed by the out-of-time-order four-point correlators. Evidently, all the action is at the boundary\footnote{Sorry, couldn't resist.}. To see how to understand these developments, particularly the Schwarzian action and its implications for chaos, in the sine-Gordon gauge, notice that the {\it pure} $AdS_{2}$ background maps to the sine-Gordon equation {\it on the infinite line}. The boundary conditions supplied at spatial infinity are famously integrable and precisely the reason we can construct its soliton spectrum via the B\"acklund transform. Cutting off the spacetime with some boundary curve has the effect of putting the sine-Gordon equation on a {\it finite, time-dependent} interval. Now, even on a static finite interval, boundary conditions that preserve integrablility are a set of measure zero, but when the endpoints of the interval are subject to external driving, the system is generically chaotic \cite{Murugan:2022kkj}, exhibiting similar characteristics to the Sinai billiard problem one dimension higher (see Figure 2).

\begin{figure}[h]
\centering
\includegraphics[width=0.35\textwidth]{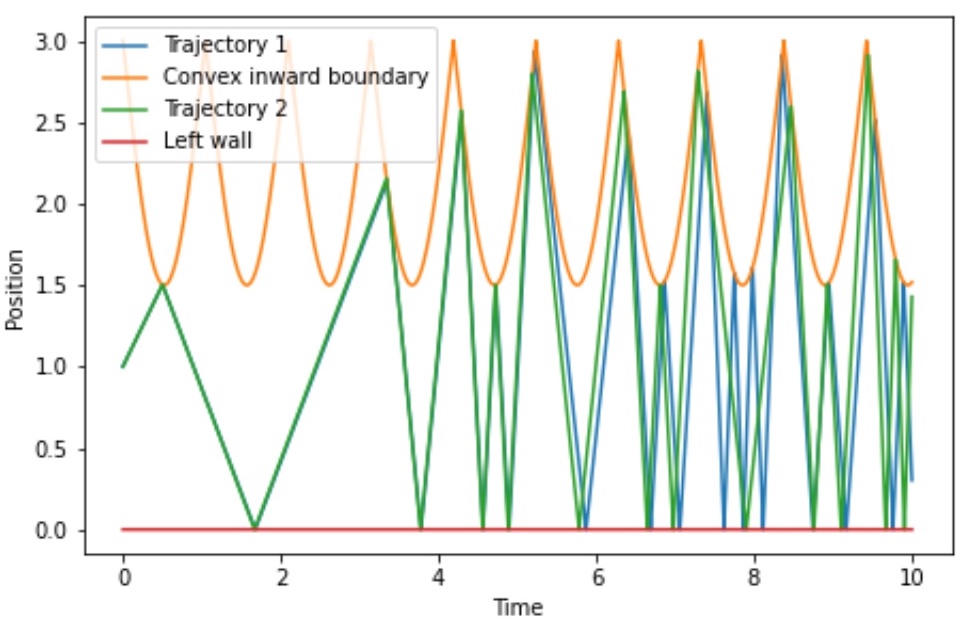}
\caption{\label{fig:TDB}A plot of the positions of two classical particles on a time-dependent finite interval, starting a distance of $10^{-3}$ away from each other with the left boundary remaining fixed and the right boundary’s position given by $L(t) = 3\left(1-\frac{1}{2}|\sin t|\right)$. The rapid divergence of trajectories is characteristic of classical chaotic behaviour.}
\end{figure}
\noindent
However, the Schwarzian limit of JT-gravity is not just mildly chaotic; as the putative dual to the SYK model, it is {\it maximally chaotic} in the sense of saturating the MSS bound on the leading Lyapunov exponent \cite{Maldacena:2015waa}. Again, chaos in the bounded sine-Gordon equation should be anticipated, but why and how such a system exhibits maximal chaos remains unclear, especially because there is no such expectation of a classical system. On the topic of boundaries, while there is an extensive literature on pseudospheres without boundary, most of which can be faithfully mapped to various solitons of the sine-Gordon equation, comparatively little is known about pseudospheres with boundaries, save that they are difficult to construct directly. On the other hand, with the map between sine-Gordon solutions and pseudospheres at hand, perhaps knowledge of sine-Gordon solitons on the finite interval, along with their various bound states may well facilitate such constructions. Our motivation for studying this problem is more utilitarian; one solution of nearly-$AdS_{2}$ gravity that has generated significant recent interest, in part because it offers an excellent laboratory for the controlled study of questions of information loss, the Hawking-Page transition, Page curves and emergent gravity, is a wormhole configuration that connects the two boundaries of the global spacetime \cite{Maldacena:2018lmt}. This Maldacena-Qi wormhole is conjectured to be dual to two coupled SYK quantum dots and it is tempting to speculate that such a ``double trumpet" solution may be constructed by gluing together two pseudospheres with boundary along the boundary, in which case the wormhole would map to a soliton bound state. Finally, there is the issue of quantization of JT gravity and the corresponding quantum sine-Gordon system, but perhaps that is a story for another Nankai symposium.

\section*{Acknowledgements}
I owe a debt of gratitude to Kayla Hopley who, in the course of her MSc research, found out about the connection between JT gravity and the sine-Gordon equation, and to Sebastian Zimper for producing Fig.2. I am also indebted to Yang-Hui He for his kind invitation to present this work at the Nankai Symposium. I regret only that I, like the other participants, could not be there in person. 

\bibliographystyle{usrt}

\begin{thebibliography}{99}

\bibitem{Chern86}
S.S. Chern and K. Tenenblat.
Studies in Applied Mathematics \textbf{74}, 55-83 (1986)

\bibitem{Teitelboim:1983ux}
C.~Teitelboim,
Phys. Lett. B \textbf{126}, 41-45 (1983)
doi:10.1016/0370-2693(83)90012-6

\bibitem{Jackiw:1984je}
R.~Jackiw,
Nucl. Phys. B \textbf{252}, 343-356 (1985)
doi:10.1016/0550-3213(85)90448-1

\bibitem{Mann:1992yq}
R.~B.~Mann and S.~F.~Ross,
Phys. Rev. D \textbf{47}, 3312-3318 (1993)
doi:10.1103/PhysRevD.47.3312
[arXiv:hep-th/9206022 [hep-th]].

\bibitem{Mann:1989gh}
R.~B.~Mann, A.~Shiekh and L.~Tarasov,
Nucl. Phys. B \textbf{341}, 134-154 (1990)
doi:10.1016/0550-3213(90)90265-F

\bibitem{Isler:1989hq}
K.~Isler and C.~A.~Trugenberger,
Phys. Rev. Lett. \textbf{63}, 834 (1989)
doi:10.1103/PhysRevLett.63.834

\bibitem{Chamseddine:1989yz}
A.~H.~Chamseddine and D.~Wyler,
Phys. Lett. B \textbf{228}, 75-78 (1989)
doi:10.1016/0370-2693(89)90528-5

\bibitem{Maldacena:2016hyu}
J.~Maldacena and D.~Stanford,
Phys. Rev. D \textbf{94}, no.10, 106002 (2016)
doi:10.1103/PhysRevD.94.106002
[arXiv:1604.07818 [hep-th]].

\bibitem{Kitaev:2017awl}
A.~Kitaev and S.~J.~Suh,
JHEP \textbf{05}, 183 (2018)
doi:10.1007/JHEP05(2018)183
[arXiv:1711.08467 [hep-th]].
\bibitem{Stanford:2019vob}
D.~Stanford and E.~Witten,
Adv. Theor. Math. Phys. \textbf{24}, no.6, 1475-1680 (2020)
doi:10.4310/ATMP.2020.v24.n6.a4
[arXiv:1907.03363 [hep-th]].

\bibitem{Penington:2019kki}
G.~Penington, S.~H.~Shenker, D.~Stanford and Z.~Yang,
[arXiv:1911.11977 [hep-th]].

\bibitem{Mertens:2018fds}
T.~G.~Mertens,
JHEP \textbf{05}, 036 (2018)
doi:10.1007/JHEP05(2018)036
[arXiv:1801.09605 [hep-th]].

\bibitem{Hollowood:2020kvk}
T.~J.~Hollowood, S.~Prem Kumar and A.~Legramandi,
J. Phys. A \textbf{53}, no.47, 475401 (2020)
doi:10.1088/1751-8121/abbc51
[arXiv:2007.04877 [hep-th]].

\bibitem{Reyes:2006rf}
E.~G.~Reyes,
J. Phys. A \textbf{39}, L55-L60 (2006)
doi:10.1088/0305-4470/39/2/L02

\bibitem{Gegenberg:1997ja}
J.~Gegenberg and G.~Kunstatter,
[arXiv:hep-th/9709183 [hep-th]].

\bibitem{Gegenberg:1997ns}
J.~Gegenberg and G.~Kunstatter,
Phys. Lett. B \textbf{413}, 274-280 (1997)
doi:10.1016/S0370-2693(97)01118-0
[arXiv:hep-th/9707181 [hep-th]].

\bibitem{Gegenberg:1998rt}
J.~Gegenberg and G.~Kunstatter,
Phys. Rev. D \textbf{58}, 124010 (1998)
doi:10.1103/PhysRevD.58.124010
[arXiv:hep-th/9807042 [hep-th]].

\bibitem{Rogers:1982qb}
C.~Rogers and W.~F.~Shadwick,

    \bibitem{Murugan:2022kki}
    J. Murugan,
    ``{\it The sine-Gordon/JT gravity connection revisited,}"
    In progress, (2022)
    
    \bibitem{Murugan:2022kkj}
    J. Murugan and S. Zimper,
    ``{\it Chaos in the sine-Gordon equation with Floquet boundaries,}"
    In progress, (2022)

\bibitem{Maldacena:2015waa}
J.~Maldacena, S.~H.~Shenker and D.~Stanford,
JHEP \textbf{08}, 106 (2016)
doi:10.1007/JHEP08(2016)106
[arXiv:1503.01409 [hep-th]].

\bibitem{Maldacena:2018lmt}
J.~Maldacena and X.~L.~Qi,
[arXiv:1804.00491 [hep-th]].
\end{thebibliography}

\end{document}